# Uniqueness of the Isotropic Frame and Usefulness of the Lorentz Transformation

Yang-Ho Choi

Department of Electrical and Electronic Engineering,
Kangwon National University,
Chuncheon, Kangwon-do 24341, South Korea

According to the postulates of the special theory of relativity (STR), physical quantities such as proper times and Doppler shifts can be obtained from any inertial frame by regarding it as isotropic. Nonetheless many inconsistencies arise from the postulates, as shown in this paper. However, there are numerous experimental results that agree with the predictions of STR. It is explained why they are accurate despite the inconsistencies. The Lorentz transformation (LT), unless subject to the postulates of STR, may be a useful method to approach physics problems. As an example to show the usefulness of LT, the problem of the generalized Sagnac effect is solved by utilizing it.

# I. INTRODUCTION

Presupposing a preferred reference frame which is isotropic so that the speed of light is invariant irrespective of its propagation direction, Mansouri and Sexl (MS) suggested a general framework for the transformation between the isotropic frame and an inertial frame [1]. Based on the MS general framework, numerous experiments associated with time dilation or the Doppler effect [2–9] have been carried out to test the validity of the special theory of relativity (STR) which postulates the principle of relativity and the constancy of the speed of light [10]. According to the postulates, any convenient inertial frame with respect to which relative velocities are readily obtained can be selected to be a preferred frame in which the speed of light is considered to be isotropic. The experimental results have shown to be in agreement with the predictions of STR, which may have led to the firm belief that STR has been experimentally verified.

In the STR, all inertial frame are equivalent and isotropic according to the postulates. Hence there are an infinite number of isotropic frames. Recently the relationship of time dilation and the formula of the Doppler effect have been derived under the uniqueness of the isotropic frame by exploiting the MS framework [11, 12]. The unique isotropic frame is the preferred reference frame. In this paper, a theory based on the uniqueness of the isotropic frame is referred to as the preferred frame theory (PFT). Using the derived results and introducing the standard synchronization into inertial frames, we examine the validity of the postulates of STR in terms of proper time and the Doppler effect, particularly focusing on the equivalence of inertial frames.

Proper times and Doppler-shifted frequencies, which are independent of the synchronization of clocks [2, 3, 11, 12], can be obtained by regarding any inertial frame as isotropic. Nonetheless, their values are demonstrated not to be uniquely determined unless the isotropic frame is unique. As a matter of fact, it is easy, as shown in Section III, to prove that the equivalence of inertial frames under the constancy of the speed of light is mathematically infeasible. If two velocities are non-collinear, the Mocanu paradox that the velocity composition law of STR is inconsistent is caused [13–15]. It results from the mathematical infeasibility. In addition, the constancy of the speed of light disagrees with the experimental result of the generalized Sagnac effect [16, 17], which clearly indicates the anisotropy of the speed of light not only in rotating frames but also in inertial frames [11, 18]. On the contrary, the PFT with a unique isotropic frame is consistent with the experimental results, including the generalized Sagnac effect, for the test of STR and it has no contradictions and no paradoxes.

Despite the infeasibility, the predictions of STR have been in agreement with numerous experimental results of time dilation or the Doppler effect. The reason for the agreement is explained. Though the postulates of STR are mathematically infeasible, the Lorentz transformation (LT) itself, if not subject to the postulates, is a very useful method, which can make mathematical manipulation easy and which needs only relative velocities between inertial frames without requiring the absolute velocities with respect to the preferred frame. We discuss the usefulness of LT, together with its limitation. It is shown that LT can exactly obtain some physical quantities, such as proper time, which are independent of the synchronization of clocks.

Neither special relativity nor general relativity could have consistently dealt with the Sagnac effect. Though a variety of explanations and analyses on the Sagnac effect are available based on these theories [19, 20], even the problem of time gap that multiple times are defined at the same place in the rotating frame has not been resolved. The generalized Sagnac effect [11, 16–18, 21] that shows the anisotropy of the speed of light in inertial frames as well may be a more perplexing conundrum to the theory of relativity based on its isotropy. We demonstrate, as an example to show the usefulness of LT, that the difference between the travel times of counter-propagating light beams in the experiment of the generalized Sagnac effect can be exactly obtained by exploiting it. Though LT can exactly discover some physical quantities, it cannot for some quantities such as spatial coordinates even if they are irrelevant to the clock synchronization. We investigate, through numerical calculations, how accurately LT can find spatial coordinate vectors.

The conclusions of the paper differ substantially from those of other articles, e.g., Refs. 15 and 22–24, which employ the Thomas rotation to explain successive transformations. According to this other point of view, no inconsistency is present in STR once the Thomas rotation is properly taken into account. It is discussed in Subsection III.2 why this standard treatment cannot be considered a satisfactory solution of the problem.

## II. GENERAL TRANSFORMATION BETWEEN INERTIAL FRAMES

This section presents a general transformation between inertial frames, together with the equations of time dilation and of the Doppler effect, based on the MS framework under the uniqueness of the isotropic frame. The preferred reference frame $S$ is isotropic so that the speed of light is a constant $c$ regardless of the propagation direction. An observer $O_l$, who is at rest in an inertial frame $S_l$, is in rectilinear motion at a velocity $v_l$ as seen in $S$. Its normalized velocity is denoted by $\beta_l$, i.e., $\beta_l = v_l/c$, which is an absolute velocity with respect to $S$ in PFT. Representing time as an imaginary number, the space-time coordinate vector of $S$ is expressed as $p = [\tau, x^T]^T$ where $\tau = ict$ with $i = (-1)^{1/2}$ and $t$ denoting time, $x$ is a spatial vector, and $T$ stands for the transpose. Similarly the coordinate vector of $S_l$ is expressed as $p_{(l)} = [\tau_{(l)}, x_{(l)}^T]^T$.

In the MS general framework, $p$ is transformed into $S_l$ in such a way that [1]

$$\tau_{(l)} = g_l \tau + i\rho_l^T x, \tag{1a}$$

$$\boldsymbol{x}_{(l)} = ib_l\tau\boldsymbol{\beta}_l + b\hat{\boldsymbol{\beta}}_l(\hat{\boldsymbol{\beta}}_l^T\boldsymbol{x}) + d_l(\boldsymbol{x} - \hat{\boldsymbol{\beta}}_l(\hat{\boldsymbol{\beta}}_l^T\boldsymbol{x})), \tag{1b}$$

where

$$g_l = a_l - b_l\boldsymbol{\varepsilon}_l^T\boldsymbol{\beta}_l, \tag{2a}$$

$$\boldsymbol{\rho}_l = (b_l - d_l)(\boldsymbol{\varepsilon}_l^T\hat{\boldsymbol{\beta}}_l)\hat{\boldsymbol{\beta}}_l + d_l\boldsymbol{\varepsilon}_l. \tag{2b}$$

In the above equations, the vectors with hats denote unit vectors. For example, given a real vector $\boldsymbol{q}$, its unit vector is denoted by $\hat{\boldsymbol{q}}$, i.e., $\hat{\boldsymbol{q}} = \boldsymbol{q}/\|\boldsymbol{q}\|$, where $\|\cdot\|$ designates the Euclidean norm. Moreover we denote its magnitude by $q$, i.e., $q = \|\boldsymbol{q}\|$. The transformation coefficients $a_l$, $b_l$ and $d_l$ can depend on $\boldsymbol{\beta}_l$, and the synchronization vector $\boldsymbol{\varepsilon}_l$ is determined by a synchronization scheme for clocks in $S_l$. The general transformation (1) can be represented in matrix form as

$$\boldsymbol{p}_{(l)} = \boldsymbol{T}_G(\boldsymbol{\beta}_l)\boldsymbol{p}, \tag{3}$$

where $\boldsymbol{T}_G(\boldsymbol{\beta}_l)$ can be written as a partitioned matrix:

$$\boldsymbol{T}_G(\boldsymbol{\beta}_l) = \begin{bmatrix} g_l & i\boldsymbol{\rho}_l^T \\ ib_l\boldsymbol{\beta}_l & \boldsymbol{M}(\boldsymbol{\beta}_l) \end{bmatrix}, \tag{4}$$

where

$$\boldsymbol{M}(\boldsymbol{\beta}_l) = (b_l - d_l)\hat{\boldsymbol{\beta}}_l\hat{\boldsymbol{\beta}}_l^T + d_l\boldsymbol{I}, \tag{5}$$

with $\boldsymbol{I}$ an identity matrix.

Using Eq. (3), one can find the transformation between arbitrary inertial frames $S_i$ and $S_j$:

$$\boldsymbol{p}_{(j)} = \boldsymbol{T}_G(\boldsymbol{\beta}_j, \boldsymbol{\beta}_i)\boldsymbol{p}_{(i)}, \tag{6}$$

where

$$\boldsymbol{T}_G(\boldsymbol{\beta}_j, \boldsymbol{\beta}_i) = \boldsymbol{T}_G(\boldsymbol{\beta}_j)\boldsymbol{T}_G^{-1}(\boldsymbol{\beta}_i). \tag{7}$$

From Eqs. (4), (5), and (7), $\boldsymbol{T}_G(\boldsymbol{\beta}_j, \boldsymbol{\beta}_i)$ is written as [12]

$$\boldsymbol{T}_G(\boldsymbol{\beta}_j, \boldsymbol{\beta}_i) = \begin{bmatrix} a_i^{-1}(g_j + \boldsymbol{\rho}_j^T\boldsymbol{\beta}_i) & -ig_ja_i^{-1}\boldsymbol{\varepsilon}_i^T + i\boldsymbol{\rho}_j^T\boldsymbol{M}'(\boldsymbol{\beta}_i) \\ ia_i^{-1}(b_j\boldsymbol{\beta}_j - \boldsymbol{M}(\boldsymbol{\beta}_j)\boldsymbol{\beta}_i) & a_i^{-1}b_j\boldsymbol{\beta}_j\boldsymbol{\varepsilon}_i^T + \boldsymbol{M}(\boldsymbol{\beta}_j)\boldsymbol{M}'(\boldsymbol{\beta}_i) \end{bmatrix}, \tag{8}$$

where

$$\boldsymbol{M}'(\boldsymbol{\beta}_l) = (\frac{1}{b_l} - \frac{1}{d_l})\hat{\boldsymbol{\beta}}_l\hat{\boldsymbol{\beta}}_l^T - \frac{1}{a_l}\boldsymbol{\beta}_l\boldsymbol{\varepsilon}_l^T + \frac{1}{d_l}\boldsymbol{I}. \tag{9}$$

Note that the transformation between $S_i$ and $S_j$ is dependent on both $\boldsymbol{\beta}_i$ and $\boldsymbol{\beta}_j$. It is obvious that $\boldsymbol{T}_G^{-1}(\boldsymbol{\beta}_j, \boldsymbol{\beta}_i) = \boldsymbol{T}_G(\boldsymbol{\beta}_i, \boldsymbol{\beta}_j)$. Given $\boldsymbol{\beta}_i$ and $\boldsymbol{\beta}_j$, the velocity $\boldsymbol{\beta}_{ji}$ in $S_i$ of an object $O_j$ is given by [2, 12]

$$\boldsymbol{\beta}_{ji} = \frac{1}{a_j \Gamma(\boldsymbol{\beta}_i, \boldsymbol{\beta}_j)} (b_i [\hat{\boldsymbol{\beta}}_i (\hat{\boldsymbol{\beta}}_i^T \boldsymbol{\beta}_j) - \boldsymbol{\beta}_i] + d_i [\boldsymbol{\beta}_j - \hat{\boldsymbol{\beta}}_i (\hat{\boldsymbol{\beta}}_i^T \boldsymbol{\beta}_j)]), \tag{10}$$

where

$$\Gamma(\boldsymbol{\beta}_i, \boldsymbol{\beta}_j) = \frac{a_i}{a_j} + \frac{b_i}{a_j} (\boldsymbol{\varepsilon}_i^T [\hat{\boldsymbol{\beta}}_i (\hat{\boldsymbol{\beta}}_i^T \boldsymbol{\beta}_j) - \boldsymbol{\beta}_i]) + \frac{d_i}{a_j} (\boldsymbol{\varepsilon}_i^T [\boldsymbol{\beta}_j - \hat{\boldsymbol{\beta}}_i (\hat{\boldsymbol{\beta}}_i^T \boldsymbol{\beta}_j)]). \tag{11}$$

It is easy to see from Eq. (10) that $\boldsymbol{\beta}_{ji}$ is reduced to $\boldsymbol{\beta}_j$ when $\boldsymbol{\beta}_i = \mathbf{0}$ so that $a_i = b_i = d_i = 1$ and $\boldsymbol{\varepsilon}_i = \mathbf{0}$. Note that the direction of $\boldsymbol{\beta}_{ji}$ is independent of $\boldsymbol{\varepsilon}_i$ and $\boldsymbol{\varepsilon}_j$ whereas its magnitude is dependent on $\boldsymbol{\varepsilon}_i$.

It is well known that proper time (PT) is independent of the synchronization of clocks. We use a subscript '∘' at PT, say $\tau_{(l)\circ}$, to distinguish it from adjusted time (AT) through a synchronization procedure. The PT interval is measured at the same position while the AT interval is the time difference between different positions. The (1,1)-entry of $\boldsymbol{T}_G(\boldsymbol{\beta}_j, \boldsymbol{\beta}_i)$ represents the time dilation factor, which makes a connection between PT and AT, and it is given by [12]

$$\boldsymbol{T}_G(\boldsymbol{\beta}_j, \boldsymbol{\beta}_i)|_{11} = \Gamma(\boldsymbol{\beta}_j, \boldsymbol{\beta}_i), \tag{12}$$

where $\boldsymbol{A}|_{mn}$, $m, n = 1, 2$, represents the $(m,n)$-entry of a partitioned matrix $\boldsymbol{A}$, i.e., $\boldsymbol{A}|_{mn} = \boldsymbol{A}_{mn}$. The differential forms for Eqs. (3) and (6) are obtained by replacing coordinate vectors with differential vectors. The differential coordinate vector of an observer $O_j$ who is at rest in $S_j$ can be represented as $d\boldsymbol{p}_{(j)} = [d\tau_{(j)\circ}, \mathbf{0}]^T$. Substituting the $d\boldsymbol{p}_{(j)}$ into the differential form of Eq. (6) with subscripts $i$ and $j$ interchanged yields

$$d\boldsymbol{p}_{(i)} = d\tau_{(j)\circ} \boldsymbol{T}_G(\boldsymbol{\beta}_i, \boldsymbol{\beta}_j)|_{1c} \tag{13a}$$
$$= d\tau_{(j)\circ} [\Gamma(\boldsymbol{\beta}_i, \boldsymbol{\beta}_j), (\boldsymbol{T}_G(\boldsymbol{\beta}_i, \boldsymbol{\beta}_j)|_{21})^T]^T, \tag{13b}$$

where $\boldsymbol{A}|_{1c}$ denotes the first column of a matrix $\boldsymbol{A}$. The differential vector $d\boldsymbol{p}_{(i)}$ can be generally expressed as

$$d\boldsymbol{p}_{(i)} = d\tau_{(i)} [1, d\boldsymbol{x}_{(i)}^T / d\tau_{(i)}]^T \tag{14a}$$
$$= d\tau_{(i)} [1, -i\boldsymbol{\beta}_{ji}^T]^T. \tag{14b}$$

Equation (14b) results from the fact that the normalized velocity of $O_j$ is $id\boldsymbol{x}_{(i)}/d\tau_{(i)} = \boldsymbol{\beta}_{ji}$ in $S_i$. Comparing Eqs. (13b) and (14b), the differential PT of $O_j$ is related to $d\tau_{(i)}$ by

$$d\tau_{(j)\circ} = \frac{d\tau_{(i)}}{\Gamma(\boldsymbol{\beta}_i, \boldsymbol{\beta}_j)}. \tag{15}$$

From Eq. (13a), $\boldsymbol{T}_G(\boldsymbol{\beta}_i, \boldsymbol{\beta}_j)|_{1c} = d\boldsymbol{p}_{(i)}/d\tau_{(j)\circ}$, which is written using Eqs. (14b) and (15) as

$$\boldsymbol{T}_G(\boldsymbol{\beta}_i, \boldsymbol{\beta}_j)|_{1c} = \Gamma(\boldsymbol{\beta}_i, \boldsymbol{\beta}_j)[1, -i\boldsymbol{\beta}_{ji}^T]^T . \tag{16}$$

The formula of the Doppler effect can be derived from Eq. (6). Consider that a plane wave of frequency $\nu_l$ propagates with a direction vector $\boldsymbol{k}_{lD} = [k_{l1}, k_{l2}, k_{l3}]^T$ in an inertial frame $S_l$. Its wave 4-vector can be written as $\boldsymbol{k}_l = [\omega_{lc}, \boldsymbol{k}_{lD}^T]^T$ where $\omega_{lc} = i\omega_l/c$ with the angular frequency $\omega_l = 2\pi\nu_l$. The wave vectors $\boldsymbol{k}_i$ and $\boldsymbol{k}_j$ are related to [12]

$$\boldsymbol{k}_j = \boldsymbol{T}_G^{-T}(\boldsymbol{\beta}_j, \boldsymbol{\beta}_i)\boldsymbol{k}_i . \tag{17}$$

The Doppler-shifted frequency $\nu_j$ as seen in $S_j$ is written as

$$\nu_j = \nu_i \Gamma(\boldsymbol{\beta}_i, \boldsymbol{\beta}_j)(1 - \beta_{ji}\cos\xi_{ji}) , \tag{18}$$

where $\xi_{ji}$ is the angle between $\boldsymbol{k}_{iD}$ and $\boldsymbol{\beta}_{ji}$ so that $\cos\xi_{ji} = \hat{\boldsymbol{\beta}}_{ji}^T\hat{\boldsymbol{k}}_{iD}$. The relationships of time dilation and of the Doppler effect are valid within the general framework of Eq. (6) regardless of the transformation coefficients and the synchronization parameters. It should be noted that the equations have been derived under the uniqueness of the isotropic frame.

### III. UNIQUENESS OF THE ISOTROPIC FRAME

The STR postulates the principle of relativity and the constancy of the speed of light. All inertial frames are equivalent and isotropic according to the postulates so that there are an infinite number of isotropic frames. Equation (3) represents a general transformation between the isotropic frame $S$ and an inertial frame. The STR and the PFT see it differently. In PFT, only $S$ is isotropic and it is the preferred reference frame. On the contrary, in STR, any inertial frame, which is regarded as isotropic, can play the role of the preferred frame. In this section, the frame that is actually isotropic is shown to be unique. However, there are many experimental results of time dilation or the relativistic Doppler effect that agree with the predictions of STR. Introducing the standard synchronization into inertial frames, we present the equations for PT and the Doppler effect under PFT and explains the reason for the agreement. Furthermore, it is shown that STR has many inconsistencies.

**1. Uniqueness of physical quantities and uniqueness of the isotropic frame**

We employ the coefficients in Eq. (3) that STR has suggested:

$$a_l = \gamma_l^{-1}, \ b_l = \gamma_l, \ d_l = 1, \tag{19}$$

where $\gamma_l = (1 - \beta_l^2)^{-1/2}$. Also, we adopt the standard synchronization so that the synchronization vector becomes $\boldsymbol{\varepsilon}_l = -\boldsymbol{\beta}_l$. Then Eqs. (2a) and (2b) are expressed as $g_l = \gamma_l$ and $\boldsymbol{\rho}_l = -\gamma\boldsymbol{\beta}_l$. In the standard synchronization, the transformation matrix $T_G(\boldsymbol{\beta}_l)$ is written as

$$T_G(\boldsymbol{\beta}_l) = \begin{bmatrix} \gamma_l & -i\gamma_l \boldsymbol{\beta}_l^T \\ i\gamma_l \boldsymbol{\beta}_l & (\gamma_l - 1)\hat{\boldsymbol{\beta}}_l \hat{\boldsymbol{\beta}}_l^T + \boldsymbol{I} \end{bmatrix}. \tag{20}$$

It is easy to see that $T_G^T(\boldsymbol{\beta}_l) T_G(\boldsymbol{\beta}_l) = \boldsymbol{I}$. Recently a coordinate transformation, called the transformation under the constant light speed (TCL), between a rotating system and the isotropic $S$ has been suggested [18]. A transformation between $S_l$ and $S$ can be derived from TCL through the limit operation of circular motion to linear motion. The derived transformation has the same form as an inertial transformation [25]. The same transformation can also be obtained through the LT of the world lines of inertial observers [26]. The inertial transformation matrix is identical to Eq. (4) with Eq. (19) and $\boldsymbol{\varepsilon}_l = \boldsymbol{0}$, which may be an exact one for inertial systems. For convenience, under the standard synchronization, we refer to Eq. (6) with Eq. (19) as an exact transformation (ET), which represents the coordinate transformation in PFT.

It is convenient to introduce a partitioned matrix to be used in place of transformation matrices:

$$\boldsymbol{A} = \begin{bmatrix} A_{11} & \boldsymbol{A}_{12} \\ \boldsymbol{A}_{21} & \boldsymbol{A}_{22} \end{bmatrix}, \tag{21}$$

where $A_{11}$ is a scalar quantity. Let $\boldsymbol{A} = \boldsymbol{T}_G(\boldsymbol{\beta}_j, \boldsymbol{\beta}_i)$. Because $T_G^{-1}(\boldsymbol{\beta}_l) = T_G^T(\boldsymbol{\beta}_l)$, $\boldsymbol{A}^{-1} = \boldsymbol{A}^T$. For simplicity, $\gamma_{ji}$ is used instead of $\Gamma(\boldsymbol{\beta}_j, \boldsymbol{\beta}_i)$. Using Eq. (16), we have [11]

$$\boldsymbol{A}_{12}^T = \boldsymbol{T}_G(\boldsymbol{\beta}_i, \boldsymbol{\beta}_j)|_{21} = -i\gamma_{ij}\boldsymbol{\beta}_{ji}, \tag{22}$$

$$A_{11} = \gamma_{ji} = \gamma_{ij}. \tag{23}$$

It follows from $\boldsymbol{A}^T \boldsymbol{A} = \boldsymbol{I}$ that $A_{11}^2 + \boldsymbol{A}_{21}^T \boldsymbol{A}_{21} = 1$. Substituting Eqs. (22) and (23) into this equation gives

$$\gamma_{ji} = (1 - \beta_{ij}^2)^{-1/2}. \tag{24}$$

Equations (23) and (24) lead to $\beta_{ji} = \beta_{ij}$. The $\boldsymbol{\beta}_{ji}$ is written from Eqs. (10) and (19) as

$$\boldsymbol{\beta}_{ji} = \gamma_j \gamma_{ij}^{-1} [\gamma_i(\hat{\boldsymbol{\beta}}_i(\hat{\boldsymbol{\beta}}_i^T \boldsymbol{\beta}_j) - \boldsymbol{\beta}_i) + \boldsymbol{\beta}_j - \hat{\boldsymbol{\beta}}_i(\hat{\boldsymbol{\beta}}_i^T \boldsymbol{\beta}_j)]. \tag{25}$$

The velocity $\boldsymbol{\beta}_{ij}$ can be found by interchanging subscripts $i$ and $j$ in Eq. (25). Though the magnitudes of $\boldsymbol{\beta}_{ji}$ and $\boldsymbol{\beta}_{ij}$ are identical, generally $\boldsymbol{\beta}_{ij} \neq -\boldsymbol{\beta}_{ji}$. If the directions of $\boldsymbol{\beta}_i$ and $\boldsymbol{\beta}_j$ are the same, however, it is readily seen from Eq. (25) that $\boldsymbol{\beta}_{ij} = -\boldsymbol{\beta}_{ji}$.

The differential PT of $O_j$ can be expressed as

$$d\tau_{(j)\circ} = d\tau_{(i)} / \gamma_{ji} \tag{26a}$$

$$= d\tau / \gamma_j. \tag{26b}$$

It is worth noting that Eq. (26a) is valid even if subscripts $i$ and $j$ are interchanged. The first row of $A$ is written from Eqs. (22) and (23) as $A|_{1r} = \gamma_{ji}[1, -i\boldsymbol{\beta}_{ji}^T]$. The motion of $O_j$ is described as $\boldsymbol{p}_{(i)} = \tau_{(i)}[1, -i\boldsymbol{\beta}_{ji}^T]^T$ in $S_i$. Recall that Eq. (3) with $l = j$ and Eq. (6) are the representations for the same $\boldsymbol{p}_{(j)}$. Equation (26a) is obtained from the substitution of the $\boldsymbol{p}_{(i)}$ into Eq. (6), and Eq. (26b) from the substitution of $\boldsymbol{p} = \tau[1, -i\boldsymbol{\beta}_j^T]^T$ into Eq. (3) with $l = j$. Note that the PT of $O_j$ is the same regardless of $\varepsilon_i$ and $\varepsilon_j$ for any $S_i$, as can be seen from Eq. (26b).

It is very important to understand the meaning of time $\tau$ in the isotropic $S$ such that differences between $\tau$ and AT in an inertial frame can be recognized. In Eq. (26b), the $d\tau$ between two different places is actually the same as the time interval at the same place. That is to say, in Fig. 1, the $d\tau_{00}$ ($= \tau_{20} - \tau_{10}$) is actually equal to the $d\tau_{21}$ ($= \tau_{22} - \tau_{11}$) because events at the same $\tau$ are actually simultaneous in $S$. Let us suppose that there exists another isotropic frame $S_0$, which is in uniform linear motion at a normalized velocity $\boldsymbol{\beta}_0$ with respect to $S$. Even if $d\tau_{(0)}$ is the time interval measured between different positions, where $\tau_{(0)}$ denotes the time coordinate in $S_0$, it is the same as the PT interval because events at the same $\tau_{(0)}$ are really simultaneous in $S_0$. The coordinates between $S$ and $S_0$ are related by LT. The velocity of $S$ relative to $S_0$ is $-\boldsymbol{\beta}_0$. The relationships of time between $S$ and $S_0$ are written as $d\tau_{(0)} = d\tau / \gamma_0$ from $\boldsymbol{p}_{(0)} = \boldsymbol{T}_G(\boldsymbol{\beta}_0)\boldsymbol{p}$ and $d\tau = d\tau_{(0)} / \gamma_0$ from $\boldsymbol{p} = \boldsymbol{T}_G(-\boldsymbol{\beta}_0)\boldsymbol{p}_{(0)}$ where $\gamma_0 = (1-\beta_0^2)^{-1/2}$. They, however, contradict each other because all the time intervals in $S_0$ and in $S$ are equivalent to the PT intervals even if they are between different positions. In fact, AT is an apparent time and PT is the correct time. In the isotropic frame, $d\tau_{21}$, $d\tau_{01}$ ($= \tau_{20} - \tau_{11}$), and $d\tau_{20}$ ($= \tau_{22} - \tau_{10}$) as well as $d\tau_{00}$ are all the correct time intervals and they are really the same.

Unless the isotropic frame is unique the Doppler effect is not uniquely determined. Suppose that a plane wave $\Psi(\tau, \boldsymbol{x}) = A \sin(\boldsymbol{k}_D^T \boldsymbol{x} + \omega_c \tau)$ of frequency $\nu$ is propagating in $S$ where $A$ is the amplitude, $\boldsymbol{k}_D$ is the direction vector, and $\omega_c = i\omega/c$ with the angular frequency $\omega = 2\pi\nu$. The Doppler-shifted frequency in $S_j$ can be expressed from Eqs. (17) – (20) as

$$\nu_j = \nu_i \gamma_{ij}(1 - \beta_{ji} \cos\xi_{ji}) \tag{27a}$$
$$= \nu \gamma_j (1 - \beta_j \cos\xi_j). \tag{27b}$$

The frequency is a physical quantity representing the reciprocal of the period of the sine wave, which is obviously irrelevant to the clock synchronization. The formula (27a) expressed as a function of the relative velocity $\boldsymbol{\beta}_{ji}$ is a different representation of the same frequency $\nu_j$ related to $\nu$ by Eq. (27b) expressed as a function of the absolute velocity $\boldsymbol{\beta}_j$. When $S_i$ employs a different synchronization $\boldsymbol{\varepsilon}_i$, rather than the standard one, $\nu_j$ is represented by a function of $\boldsymbol{\varepsilon}_i$ as in Eq. (18), but it is the same regardless of $\boldsymbol{\varepsilon}_i$, as shown in Eq. (27b).

If inertial frames are equivalent, however, $\nu_j$ does not have a unique value. Consider another isotropic frame $S_0$ as above. The wave vector and the frequency in $S_0$ of the plane wave are denoted by $\boldsymbol{k}_0$ and $\nu_0$. The frequency $\nu_0$ can be written as $\nu_0 = \nu \gamma_0 (1 - \beta_0 \cos \xi_0)$ according to Eq. (27a). Because $S_0$ is considered isotropic, $\boldsymbol{k}_0$ and $\boldsymbol{k}_j$ are related by $\boldsymbol{k}_j = \boldsymbol{T}_G^{-T}(\boldsymbol{\beta}_{j0})\boldsymbol{k}_0$, which leads the Doppler-shifted frequency in $S_j$ to be given by $\nu_{j0} = \nu_0 \gamma_{j0}(1 - \beta_{j0} \cos \xi_{j0})$ where $\boldsymbol{\beta}_{j0}$ is the velocity of $S_j$ relative to $S_0$. In general, the $\nu_j$ of Eq. (27) will not be equal to the $\nu_{j0}$. It is not uniquely given. The values of proper time and frequency are unique. They have unique values under the uniqueness of the isotropic frame.

## 2. Inconsistencies in STR

In STR, coordinates between inertial frames are related by LT. A coordinate vector $\boldsymbol{p}_{(i)}$ in $S_i$ is Lorentz-transformed into $S_j$ and then the resultant vector $\overline{\boldsymbol{p}}_{(j)} = [\overline{\tau}_{(j)}, \overline{\boldsymbol{x}}_{(j)}^T]^T$ is given by

$$\overline{\boldsymbol{p}}_{(j)} = \boldsymbol{T}_L(\boldsymbol{\beta}_{ji}) \boldsymbol{p}_{(i)}, \tag{28}$$

where

$$\boldsymbol{T}_L(\boldsymbol{\beta}_{ji}) = \begin{bmatrix} \gamma_{ji} & -i\gamma_{ji}\boldsymbol{\beta}_{ji}^T \\ i\gamma_{ji}\boldsymbol{\beta}_{ji} & (\gamma_{ji} - 1)\hat{\boldsymbol{\beta}}_{ji}\hat{\boldsymbol{\beta}}_{ji}^T + \boldsymbol{I} \end{bmatrix}. \tag{29}$$

Equation (28) represents a generalized LT that can be applied to arbitrary $\boldsymbol{\beta}_{ji}$ regardless of its direction. If $\boldsymbol{\beta}_i = \boldsymbol{0}$ so that $\boldsymbol{\beta}_{ji} = \boldsymbol{\beta}_j$, the matrix $\boldsymbol{T}_L(\boldsymbol{\beta}_{ji})$ becomes equal to Eq. (20) with $\boldsymbol{\beta}_l = \boldsymbol{\beta}_j$ and the coordinate vector in STR is the same as the exact one in PFT, i.e., $\overline{\boldsymbol{p}}_{(j)} = \boldsymbol{p}_{(j)}$. The differential PT of $O_j$ when calculated from Eq. (28) is written as

$$d\overline{\tau}_{(j)} = d\tau_{(i)} / \gamma_{ji}. \tag{30}$$

If the coordinate transformation matrix is $\boldsymbol{T}_L(\boldsymbol{\beta}_{ji})$, the formula for the Doppler effect is given by $\overline{\boldsymbol{k}}_j = \boldsymbol{T}_L^{-T}(\boldsymbol{\beta}_{ji})\boldsymbol{k}_i$ and the Doppler-shifted frequency $\overline{\nu}_j$ is expressed as

$$\overline{v}_j = v_i \gamma_{ji}(1 - \beta_{ji} \cos \xi_{ji}). \tag{31}$$

Note that $d\overline{\tau}_{(j)}$ and $\overline{v}_j$ are equal to the exact ones $d\tau_{(j)\circ}$ and $v_j$ though the LT is not exact.

When the correct coordinate vector $\boldsymbol{p}_{(i)}$ is directly transformed into $S_j$ as Eq. (28), the time component $\overline{\tau}_{(j)}$ of $\overline{\boldsymbol{p}}_{(j)}$ is the same as that of $\boldsymbol{p}_{(j)}$ given by Eq. (6) but their spatial vectors are different. The transformation (28) does not provide correct coordinates. Additionally, if the coordinate transformation is made by way of another inertial frame, say $S_k$, the resultant coordinate vector, depending on the intermediate frame $S_k$, is not uniquely determined. Suppose that $\boldsymbol{p}_{(i)}$, $\boldsymbol{\beta}_{ki}$, and $\boldsymbol{\beta}_{jk}$ are given. Then STR calculates the coordinate vector in $S_j$ in such a way that

$$\overline{\boldsymbol{p}}_{(j)/k} = \boldsymbol{T}_L(\boldsymbol{\beta}_{jk}, \boldsymbol{\beta}_{ki})\boldsymbol{p}_{(i)}, \tag{32}$$

where $\boldsymbol{T}_L(\boldsymbol{\beta}_{jk}, \boldsymbol{\beta}_{ki}) = \boldsymbol{T}_L(\boldsymbol{\beta}_{jk})\boldsymbol{T}_L(\boldsymbol{\beta}_{ki})$. The vector $\overline{\boldsymbol{p}}_{(j)/k}$ is other than the $\overline{\boldsymbol{p}}_{(j)}$ of Eq. (28). It varies with the selection of the intermediate frame. The one-way speed of light has been known to be empirically inaccessible, which leaves room for the conventionality of simultaneity [6, 20, 27] to play. Spatial coordinates are irrelevant to the clock synchronization. The $\boldsymbol{p}_{(j)}$ of Eq. (6) is also written as $\boldsymbol{p}_{(j)} = \boldsymbol{T}_G(\boldsymbol{\beta}_j)\boldsymbol{p}$ and its spatial vector, though appearing to depend on $\boldsymbol{\varepsilon}_i$ and $\boldsymbol{\varepsilon}_j$ in Eq. (6), is independent of the synchronizations. However, if $S_i$ and $S_k$ employ synchronization vectors $\boldsymbol{\varepsilon}_i$ and $\boldsymbol{\varepsilon}_k$, the spatial vector of $\overline{\boldsymbol{p}}_{(j)/k}$ is dependent on them.

These inconsistencies result from the non-equality

$$\boldsymbol{T}_L(\boldsymbol{\beta}_{ji}) \neq \boldsymbol{T}_G(\boldsymbol{\beta}_j, \boldsymbol{\beta}_i), \tag{33}$$

though the standard synchronization is adopted into both $S_i$ and $S_j$. Any inertial frame can play the role of the preferred frame in STR. If $S_k$ is selected as the preferred frame, accordingly the equality $\boldsymbol{T}_L(\boldsymbol{\beta}_{ji}) = \boldsymbol{T}_L(\boldsymbol{\beta}_{jk})\boldsymbol{T}_L^{-1}(\boldsymbol{\beta}_{ik})$ should be satisfied. The frames $S_i$, $S_j$, and $S_k$ are arbitrary and it should be satisfied for every $\boldsymbol{\beta}_{ji}$, $\boldsymbol{\beta}_{jk}$, and $\boldsymbol{\beta}_{ik}$. However, the equality is mathematically infeasible, which means that the equivalence of inertial frames under the constancy of the speed of light is mathematically infeasible. It is easy to see the mathematical infeasibility unless we adhere to the belief that STR must be correct. However, the firm belief in the correctness may have hindered many people from seeing the infeasibility that can be discovered easily. Let us illustrate the Mocanu paradox [13–15].

When $S_i$ and $S_j$ are connected to an arbitrary intermediate frame $S_l$ by velocities $\boldsymbol{\beta}_{li}$ and $\boldsymbol{\beta}_{jl}$, the velocity of $S_j$ relative to $S_i$ in STR is calculated, according to its velocity composition

law [13–15], as

$$\bar{\beta}_{ji/l} = \beta_{li} \oplus \beta_{jl}, \tag{34a}$$

where

$$\beta_{li} \oplus \beta_{jl} = \frac{1}{1+\beta_{ki}^T \beta_{jl}} \left[ \beta_{li} + \frac{\beta_{jl}}{\gamma_{li}} + \frac{\gamma_{li}(\beta_{li}^T \beta_{jl})\beta_{li}}{1+\gamma_{li}} \right]. \tag{34b}$$

It is easy to see that Eq. (25) is represented as $\beta_{ji} = (-\beta_i) \oplus \beta_j$. If velocities are non-collinear, the composition law is not consistent with the postulation of STR. In general $\bar{\beta}_{ij/l} \neq -\bar{\beta}_{ji/l}$, though $\bar{\beta}_{ij/l}$ should be equal to $-\bar{\beta}_{ji/l}$ in accordance with the principle of relativity, and the Mocanu paradox is raised.

The paradox has been explained by resorting to the Thomas rotation [14, 15, 22–24, 28]. Let $S_l = S_k$ and the transformation from $S_i$ to $S_j$ is made through an inertial frame $S_k$. The successive LT of $T_L(\beta_{jk}, \beta_{ki})$ is not reduced to the LT for $\bar{\beta}_{ji/k}$, namely

$$T_L(\bar{\beta}_{ji/k}) \neq T_L(\beta_{jk}, \beta_{ki}). \tag{35}$$

For the resolution of the problem of the non-equality, a Thomas rotation is employed such that

$$R(\beta_{jk}, \beta_{ki})T_L(\bar{\beta}_{ji/k}) = T_L(\beta_{jk}, \beta_{ki}), \tag{36}$$

where the spatial rotation matrix $R(\beta_{jk}, \beta_{ki})$ is determined to satisfy the equation. The introduction of the spatial rotation may shed light on the non-equality of Eq. (35). However, the rotation matrix, depending on the intermediate frame $S_k$, is not uniquely given and $T_L(\beta_{jk}, \beta_{ki})$ is not identical to $T_G(\beta_j, \beta_i)$ so that the coordinate vector $\bar{p}_{(j)/k}$ from the former is other than $p_{(j)}$ from the latter, even their time components being different. Moreover, when the coordinates of $S_i$ are transformed into $S_j$, the resulting coordinates must always be the same irrespective of whether they are obtained directly from $S_i$ to $S_j$ or via $S_k$ or via another inertial frame $S_m$. Even if the rotation is introduced, on the contrary, the transformed coordinates do vary with the intermediate because

$$T_L(\beta_{jk}, \beta_{ki}) \neq T_L(\beta_{jm}, \beta_{mi}). \tag{37}$$

Inconsistencies in STR remain regardless of the introduction of the rotation.

Let us investigate the velocity of $S_j$ with respect to $S_i$ in the presence of two additional inertial frames $S_k$ and $S_m$ to see the inconsistency of STR more clearly. According to the velocity composition law, $\beta_{jk} = \beta_{mk} \oplus \beta_{jm}$ and $\beta_{mi} = \beta_{ki} \oplus \beta_{mk}$. Using these and Eq. (34a), we have

$$\overline{\beta}_{ji/k} = \beta_{ki} \oplus (\beta_{mk} \oplus \beta_{jm}), \tag{38a}$$

$$\overline{\beta}_{ji/m} = (\beta_{ki} \oplus \beta_{mk}) \oplus \beta_{jm}. \tag{38b}$$

The PT and the Doppler shift in $S_j$ can be obtained from relative velocities, as shown in Eqs. (26a) and (27a). The composition operation is not associative [15]. Thus $\overline{\beta}_{ji/k} \neq \overline{\beta}_{ji/m}$, which indicates that the PT and the Doppler shift also vary with the intermediate frame so that they are not uniquely given. In fact, they have an infinite number of values as there are an infinite number of intermediate frames. Moreover neither $\overline{\beta}_{ji/k}$ nor $\overline{\beta}_{ji/m}$ corresponds to $\beta_{ji}$ which is, depending only on the absolute velocities as shown in Eq. (25) (or Eq. (10)), independent of intermediate frames. Even if the correct velocity $\beta_{ji}$ is known, $T_L(\beta_{ji})$ is not equal to $T_G(\beta_j, \beta_i)$.

The spatial rotation was introduced only to explain the fact that $T_L(\beta_{jk}, \beta_{ki})$ is not reduced to $T_L(\overline{\beta}_{ji/k})$. It can explain neither the non-equality of Eq. (37) nor Eq. (33), which causes inconsistencies and contradictions in STR such as the non-uniqueness of PT, Doppler effect, and spatial coordinates. The inconsistencies and the contradictions cannot be resolved by resorting to the rotation at all. On the other hand, there are not any inconsistencies or any paradoxes in PFT that retains the equality

$$T_G(\beta_j, \beta_i) = T_G(\beta_j, \beta_k) T_G(\beta_k, \beta_i) \tag{39}$$

for every $\beta_i$, $\beta_j$, and $\beta_k$. The relationship of Eq. (39) leads to the PT, the Doppler shift, and the spatial vector having unique values from frame to frame irrespective of the clock synchronizations.

## 3. Empirical evidences

There are many experimental results of time dilation or the Doppler effect, which have been known to agree very well with the predictions of STR [2–9]. It is usually stated that STR has been experimentally confirmed. As a matter of fact, the experimental results are the evidences for PFT, rather than STR, because the equivalence of inertial frames under the light speed constancy is mathematically infeasible. Even the analyses of experimental results have often been made based on PFT under a unique isotropic frame [2, 3].

One way to prove STR and to disprove PFT is to experimentally show the isotropy of the one-way speed of light [3, 4]. To the end, the variation of the Doppler effect can be measured as it is a function of the one-way speed. The underlying idea of isotropy tests seems to be that if PFT is correct, the Doppler-shifted frequency will vary according to the velocity of a measuring apparatus with respect to the preferred frame $S$. In the isotropy test, the apparatus is placed on a rotating platform or the surface of the Earth. For example, when a test is performed to observe possible sidereal modulation on the Doppler shift due to the Earth's rotation [4], our Solar System can be considered to belong to an inertial frame $S_i$. If the velocity $\boldsymbol{\beta}_j$ of the apparatus periodically changes with respect to $S$ periodic changes recorded on the Doppler shift can be observed according to Eq. (27b). However, if there are no relative changes between the velocity $\boldsymbol{\beta}_{ji}$ and the wave vector $\boldsymbol{k}_{iD}$, when seen in the frame $S_i$ disguised as an isotropic frame, the Doppler shift does not vary, as shown in Eq. (27a), even though they change in $S$.

Sometimes STR and PFT have been considered empirically indistinguishable. In Ref. 29, the inability of detection of an ether wind is described as a conspiracy of nature. It is the equalities between Eqs. (26a) and (26b) and between Eqs. (27a) and (27b) that what nature conspires is. In the article of MS [1], "a theory maintaining absolute simultaneity is equivalent to special relativity." The theory maintaining absolute simultaneity seems to be equivalent to PFT. However, MS draw the conclusion from the investigation only under a unique isotropic system. They have investigated clock synchronizations between the isotropic frame $S$ and an inertial frame and discovered that the length contractions in STR and PFT are equal. The left side of Eq. (33) should be equal to the right side if STR is equivalent to PFT.

Equations (26) and (27) have been derived under the uniqueness of the isotropic system. Nonetheless, PTs and Doppler shifts, according to Eqs. (26a) and (27a), are exactly obtained from any inertial frame $S_i$, which has been verified through numerously repeated experiments. It is stated in Ref. 27 that "the 'suitable IRF' is nothing but 'the most convenient' one. Let us mention some examples actually considered in Selleri's papers: if the rotating reference frame is the Earth, the 'suitable IRF' is the Earth-Centered IRF $S_T$; if the rotating reference frame is a beam of relativistic muons in a storage ring, the 'suitable IRF' is the laboratory frame, at rest on the Earth … if the rotating reference frame is a rotating platform, the 'suitable IRF' is the central IRF. Last but not least, it seems surprising that, in all these examples, the only serious candidate to the role of ether rest frame, namely the IRF in which the cosmic background radiation is isotropic, keeps playing no role at all. So it should be realized (or at least suspected) that the 'ether rest frame' is nothing but a misleading expression which can be used for every useful IRF, contrary to a Lorentz-like approach and according to a relativistic approach." In the quoted sentences double quotation marks were changed to single quotation marks. The fact that PTs and Doppler shifts can be obtained from any inertial frame disguised as a preferred one is well described. If $\boldsymbol{\beta}_{ji}$ is known, under the standard synchronization, we can select any $S_i$. The PFT also can use the convenient $S_i$ as the suitable IRF (inertial reference frame), introducing the standard synchronization, if $\boldsymbol{\beta}_{ji}$ is readily found in $S_i$. Then PTs and Doppler shifts can be exactly found through LT under the uniqueness of the isotropic frame, as explained in Section IV. The reason for "keeps playing no role at all" is the equalities in Eqs. (26) and (27).

Recall $A = T_G(\beta_j, \beta_i)$. The first row of $A$ is $A|_{1r} = \gamma_{ji}[1, -i\beta_{ji}^T]$, as explained in Subsection III.1. Though the equivalence of inertial frames under the light speed constancy is mathematically infeasible, STR satisfies Eqs. (26a) and (27a), as shown in Eqs. (30) and (31), regardless of the directions of $\beta_i$ and $\beta_j$ for any $S_i$ if $\beta_{ji}$ is correctly given. The reason why it satisfies them is because the first rows of $T_L(\beta_{ji})$ and $T_G(\beta_j, \beta_i)$ are identical. However, STR cannot provide the correct predictions if the transformation from $S_i$ to $S_j$ is made via an intermediate frame, say $S_k$. In that case, it remains the same in PFT, as shown in Eq. (39), but $T_L(\beta_{ji}) \neq T_L(\beta_{jk}, \beta_{ki})$ in STR. The first rows of $T_G(\beta_j, \beta_i)$ and $T_L(\beta_{jk}, \beta_{ki})$ are different because $\beta_{ji} \neq \overline{\beta}_{ji/k}$. As a result, the predictions of STR become incorrect.

Though empirical differences between STR and PFT may not be seen from PTs, Doppler shifts, and two-way speeds of light, we can see them from the one-way speed of light. Of course, we may be unable to carry out experiments to directly measure the one-way speed because of the problem of the synchronization between distant clocks. However, we can know the speed by analyzing some empirical results related to it. The global positioning system (GPS) has been known to provide very accurate position and time information by compensating for relativistic effects [20, 30]. The GPS positioning implicitly needs the one-way speed of light to find the information. Thus, it can be calculated from the GPS navigation equations, which shows that the speed of light is anisotropic in the Earth frame [31]. It is known that the so-called Sagnac correction due to the Earth's rotation should be made for accurate information. In fact, the correction is needed on account of the anisotropy of the light speed. The Sagnac effect has been observed in inertial frames as well as in rotating frames [16, 17]. The generalized Sagnac effect that involves both linear and circular motions can be analyzed based on TCL [18]. We can also make an analysis of the effect by using the MS framework [11], taking account of the motion of the laboratory frame. These theoretical analyses, which correspond with the experimental results, indicate that the one-way speed of light is anisotropic in inertial frames as well as in rotating frames. The inertial transformation [18, 25, 26], which is consistent with PFT, shows the anisotropy of the speed of light as well.

## IV. LORENTZ TRANSFORMATION AND ITS USEFULNESS

As explained in Section III, the isotropic frame is unique. Exact physical quantities can be obtained through the ET. However, though the ET requires the absolute velocities $\beta_i$ and $\beta_j$ with respect to the preferred frame $S$, they are unknown. On the contrary, the LT needs only the relative velocity $\boldsymbol{\beta}_{ji}$ to make the transformation from $S_i$ to $S_j$. Besides, it can allow us to readily approach physics problems by easy mathematical manipulation. The LT should be used, but without the postulates and with its exact meaning and its limitation. With this viewpoint, we investigate physical quantities that can be exactly obtained by LT. The spatial vector from LT is not exact. We examine how accurately LT can provide it, through numerical calculations.

**1. Exact physical quantities obtainable**

Though generally $\boldsymbol{T}_L(\boldsymbol{\beta}_{ji})$ is not identical to $\boldsymbol{T}_G(\boldsymbol{\beta}_j, \boldsymbol{\beta}_i)$, they become equal if $\boldsymbol{\beta}_i$ and $\boldsymbol{\beta}_j$ are collinear. Recall $\boldsymbol{A} = \boldsymbol{T}_G(\boldsymbol{\beta}_j, \boldsymbol{\beta}_i)$. The matrix $\boldsymbol{A}_{22}$ is expressed from Eqs. (5), (8), (9), and (19) as

$$\boldsymbol{A}_{22} = [\varsigma(\gamma_j\gamma_i - \gamma_j - \gamma_i + 1) - \gamma_j\gamma_i\beta_j\beta_i]\hat{\boldsymbol{\beta}}_j\hat{\boldsymbol{\beta}}_i^T + (\gamma_j - 1)\hat{\boldsymbol{\beta}}_j\hat{\boldsymbol{\beta}}_j^T + (\gamma_i - 1)\hat{\boldsymbol{\beta}}_i\hat{\boldsymbol{\beta}}_i^T + \boldsymbol{I}, \quad (40)$$

where $\varsigma = \hat{\boldsymbol{\beta}}_j^T\hat{\boldsymbol{\beta}}_i$. Suppose that the directions of $\boldsymbol{\beta}_i$ and $\boldsymbol{\beta}_j$ are the same so that $\varsigma = 1$. Then, $\boldsymbol{\beta}_{ij} = -\boldsymbol{\beta}_{ji}$, and the first column of $\boldsymbol{T}_L(\boldsymbol{\beta}_{ji})$ is the same as that of $\boldsymbol{T}_G(\boldsymbol{\beta}_j, \boldsymbol{\beta}_i)$. The former becomes equal to the latter if the $(2,2)$-entries of their partitioned matrices equal. The $\gamma_{ji}$ and the $\boldsymbol{T}_G(\boldsymbol{\beta}_j, \boldsymbol{\beta}_i)|_{22}$ are calculated from Eqs. (11), (19), and (40) as

$$\gamma_{ji} = \gamma_i\gamma_j(1 - \beta_i\beta_j), \quad (41)$$

$$\boldsymbol{T}_G(\boldsymbol{\beta}_j, \boldsymbol{\beta}_i)|_{22} = (\gamma_{ji} - 1)\hat{\boldsymbol{\beta}}_j\hat{\boldsymbol{\beta}}_j^T + \boldsymbol{I}. \quad (42)$$

As $\boldsymbol{\beta}_{ji}$ is a linear combination of $\boldsymbol{\beta}_i$ and $\boldsymbol{\beta}_j$, it is obvious that

$$\hat{\boldsymbol{\beta}}_{ji}\hat{\boldsymbol{\beta}}_{ji}^T = \hat{\boldsymbol{\beta}}_j\hat{\boldsymbol{\beta}}_j^T. \quad (43)$$

From Eqs. (29), (42), and (43), $\boldsymbol{T}_G(\boldsymbol{\beta}_j, \boldsymbol{\beta}_i)|_{22} = \boldsymbol{T}_L(\boldsymbol{\beta}_{ji})|_{22}$. Next, consider a case that $\hat{\boldsymbol{\beta}}_i = -\hat{\boldsymbol{\beta}}_j$. In that case, one can readily see that $\boldsymbol{\beta}_{ij} = -\boldsymbol{\beta}_{ji}$, $\gamma_{ji} = \gamma_i\gamma_j(1 + \beta_i\beta_j)$, and $\boldsymbol{T}_G(\boldsymbol{\beta}_j, \boldsymbol{\beta}_i)|_{22}$ is represented as Eq. (42). Equation (43) is also valid for $\hat{\boldsymbol{\beta}}_i = -\hat{\boldsymbol{\beta}}_j$, and $\boldsymbol{T}_L(\boldsymbol{\beta}_{ji})$ is identical with $\boldsymbol{T}_G(\boldsymbol{\beta}_j, \boldsymbol{\beta}_i)$. Similarly it can be shown that $\boldsymbol{T}_L(\boldsymbol{\beta}_{ji}) = \boldsymbol{T}_L(\boldsymbol{\beta}_{jk}, \boldsymbol{\beta}_{ki})$ if $\boldsymbol{\beta}_{jk}$ and $\boldsymbol{\beta}_{ki}$ are collinear. Usually the LT has been used under the collinear condition, which may have led to the misunderstanding that STR is consistent despite $\boldsymbol{T}_L(\boldsymbol{\beta}_{ji}) \neq \boldsymbol{T}_L(\boldsymbol{\beta}_{jk}, \boldsymbol{\beta}_{ki})$ generally.

Recall $\boldsymbol{T}_G^T(\boldsymbol{\beta}_j, \boldsymbol{\beta}_i) = \boldsymbol{T}_G^{-1}(\boldsymbol{\beta}_j, \boldsymbol{\beta}_i)$. The invariant interval is retained in $\boldsymbol{T}_G(\boldsymbol{\beta}_j, \boldsymbol{\beta}_i)$ as well as $\boldsymbol{T}_L(\boldsymbol{\beta}_{ji})$. The coordinate vector $\bar{\boldsymbol{p}}_{(j)}$ is related to $\boldsymbol{p}_{(i)}$ by Eq. (28). Because $\bar{\tau}_{(j)} = \tau_{(j)}$, $\bar{l}_{(j)} = l_{(j)}$ where $\bar{l}_{(j)} = \|\bar{\boldsymbol{x}}_{(j)}\|$ and $l_{(j)} = \|\boldsymbol{x}_{(j)}\|$. The same spatial length as the exact one is attained from the

LT. On the other hand, in general, the entries of $\bar{\boldsymbol{x}}_{(j)}$ are different from those of $\boldsymbol{x}_{(j)}$, which are independent of synchronization schemes. As $d\bar{\tau}_{(j)} = d\tau_{(j)}$ and $d\bar{l}_{(j)} = dl_{(j)}$, where $d\bar{l}_{(j)} = \|d\bar{\boldsymbol{x}}_{(j)}\|$ and $dl_{(j)} = \|d\boldsymbol{x}_{(j)}\|$, one can also discover exact physical quantities by combining $d\bar{\tau}_{(j)}$ and $d\bar{l}_{(j)}$ like $d\bar{l}_{(j)}/d\bar{t}_{(j)}$.

As an example of discovering exact quantities from LT, let us deal with the Hafele–Keating (HK) experiment [32]. In the HK experiment, three observers $\widetilde{O}_1$, $\widetilde{O}_2$, and $\widetilde{O}_3$ are involved. One observer $\widetilde{O}_1$ is at rest on the Earth while the others $\widetilde{O}_2$ and $\widetilde{O}_3$ fly around the world in the direction of rotation of the Earth and in the opposite direction. The exact PTs of the observers can be found even if the Solar System is considered isotropic by introducing the standard synchronization. The observer $\widetilde{O}_m$, $m = 1, \cdots, 3$, is assumed to move at a constant speed $\beta_m$ in the standard-synchronized frame $S_i$ of the Solar System. They are in circular motion. It is necessary to approximate circular motions as linear motions such that LT can handle them. Figure 2 shows an approximation to a circle by $n$ line segments. As $n$ tends to infinity, the linearized shape approaches a circle. The line segments belong to different inertial frames because the directions of the velocities are different though their magnitudes are the same. In Fig. 2, when an observer $\widetilde{O}_m$ belongs to an inertial frame $S_j$, the travel time as seen in $S_i$, which is AT, is $\Delta t_{mj} = cdl_j/\beta_m$ where $dl_j$ is the length of the segment $p_{j-1}p_j$. The overall travel time by the clock of $\widetilde{O}_m$, which is PT, is given by $\Delta t_{m\circ} = cl_m/\gamma_m\beta_m$ where $\gamma_m = (1-\beta_m^2)^{-1/2}$ and $l_m$ is the travel distance in $S_i$.

The generalized Sagnac effect, in which fringe shifts are observed by the difference between the travel times of two counter-propagating light beams traversing an optical fiber loop, can also be solved by using LT. Special relativity and/or general relativity cannot consistently explain the Sagnac effect even if it does not involve uniform linear motion [19, 20]. Even the speed of light in a rotating frame has remained unsolved under the theory of relativity based on the constancy of the speed of light. However, as exact spatial lengths and PTs are calculated from LT, the exact time difference can be obtained.

In the experiment of the generalized Sagnac effect [16, 17], an optical fiber loop rotates together with the light source and detector at a normalized speed of $\beta$ with respect to the laboratory frame, which can be considered to belong to an inertial frame $S_i$. Two light beams $b_+$ and $b_-$ emitted from the source at the same time travel around the optical fiber loop in opposition directions. The fiber loop that has an arbitrary shape can be approximately represented by $n$ line segments similarly to

Fig. 2. Consider that a light beam travels to the *j*th segment which belongs to an inertial frame $S_j$ moving at a velocity $\boldsymbol{\beta}_{ji}$ relative to $S_i$. Even though LT is employed for the transformation between $S_i$ and $S_j$, the speed of light with respect to PT in $S_j$ is equal to the one from ET [11], as shown in the following.

Let $\boldsymbol{A} = \boldsymbol{T}_L(\boldsymbol{\beta}_{ji})$. Using Eqs. (21) and (28), the differential spatial vector $d\bar{\boldsymbol{x}}_{(j)}$ is written as

$$d\bar{\boldsymbol{x}}_{(j)} = d\tau_{(i)}(\boldsymbol{A}_{21} + \boldsymbol{A}_{22}\boldsymbol{c}_{i\tau}), \tag{44}$$

where $\boldsymbol{c}_{i\tau} = d\boldsymbol{x}_{(i)}/d\tau_{(i)}$. As $\boldsymbol{A}^T\boldsymbol{A} = \boldsymbol{I}$, it follows that

$$\boldsymbol{A}_{21}^T\boldsymbol{A}_{21} = 1 - A_{11}^2, \quad \boldsymbol{A}_{21}^T\boldsymbol{A}_{22} = -A_{11}\boldsymbol{A}_{12}, \quad \boldsymbol{A}_{22}^T\boldsymbol{A}_{22} = \boldsymbol{I} - \boldsymbol{A}_{12}^T\boldsymbol{A}_{12}. \tag{45}$$

The squared magnitude of $d\bar{\boldsymbol{x}}_{(j)}$ is given from Eqs. (44) and (45) by

$$\|d\bar{\boldsymbol{x}}_{(j)}\|^2 = d\tau_{(i)}^2[(1 - A_{11}^2) - 2A_{11}\boldsymbol{A}_{12}\boldsymbol{c}_{i\tau} + \boldsymbol{c}_{i\tau}^T(\boldsymbol{I} - \boldsymbol{A}_{12}^T\boldsymbol{A}_{12})\boldsymbol{c}_{i\tau}]. \tag{46}$$

Because $S_i$ is standard-synchronized, for a light beam $\|\boldsymbol{c}_{i\tau}\|^2 = -1$. Using Eq. (46) and recalling that $A_{11} = \gamma_{ji}$ and $\boldsymbol{A}_{12} = -i\gamma_{ji}\boldsymbol{\beta}_{ji}^T$, we have

$$d\bar{l}_j = -\gamma_{ji}d\tau_{(i)}(i + \boldsymbol{\beta}_{ji}^T\boldsymbol{c}_{i\tau}). \tag{47}$$

Clearly $d\bar{\tau}_{(j)\circ}$ and $d\tau_{(i)}$ are related by $d\bar{\tau}_{(j)\circ} = d\tau_{(i)}/\gamma_{ji}$ in LT. The speed of light with respect to the PT of an observer $O_j$ at rest in $S_j$ is calculated as

$$\bar{c}_j = \frac{d\bar{l}_j}{d\bar{t}_{(j)\circ}} = \gamma_{ji}^2 c(1 - \boldsymbol{\beta}_{ji}^T\hat{\boldsymbol{c}}_i). \tag{48}$$

The speed of light, $\bar{c}_j$, calculated from LT is the same as that from ET. Equation (48) shows that $\bar{c}_j$ depends on the propagation direction, which indicates the anisotropy of the speed of light in inertial frames.

The speeds of the co-rotating $b_+$ and counter-rotating $b_-$ are written as $\bar{c}_{j\pm} = \gamma_{ji}^2 c(1 - \boldsymbol{\beta}_{ji}^T\hat{\boldsymbol{c}}_{i\pm})$ where $\bar{c}_{j\pm}$ ($\boldsymbol{c}_{i\pm}$) are the speeds (velocities) of $b_\pm$ in $S_j(S_i)$. The elapsed times during the travels of $b_\pm$ in $S_j$ are calculated as

$$d\bar{t}_{(j)\circ\pm} = \frac{d\bar{l}_j}{\bar{c}_{j\pm}} = \frac{d\bar{l}_j}{\gamma_{ji}^2 c(1 - \boldsymbol{\beta}_{ji}^T\hat{\boldsymbol{c}}_{i\pm})}, \tag{49}$$

where $d\bar{t}_{(j)\circ\pm}$ are the elapsed PTs when $b_\pm$ traverse the respective paths. Because $b_+$ and $b_-$ travel in opposite directions, $\boldsymbol{c}_{i-} = -\boldsymbol{c}_{i+}$. The time difference in $S_j$ is given by

$$\Delta \bar{t}_j = d\bar{t}_{(j)\circ+} - d\bar{t}_{(j)\circ-} = \frac{2 d\bar{l}_j \boldsymbol{\beta}_{ji}^T \hat{\boldsymbol{c}}_{i+}}{\gamma_{ji}^2 c[1-(\boldsymbol{\beta}_{ji}^T \hat{\boldsymbol{c}}_{i+})^2]}. \tag{50}$$

As the direction of $\boldsymbol{c}_{i+}$ is identical with that of $\boldsymbol{\beta}_{ji}$, $\boldsymbol{\beta}_{ji}^T \hat{\boldsymbol{c}}_{i+} = \beta_{ji}$ and the denominator in Eq. (49) is reduced to $c$. The differential segments that compose the fiber loop move at the same speed, and thus all $\beta_{ji}$ are equal to $\beta$. Then

$$\Delta \bar{t}_j = \frac{2 d\bar{l}_j \beta}{c}. \tag{51}$$

The total time difference $\Delta t_F$ observed at the detector is written as

$$\Delta \bar{t}_F = \lim_{n \to \infty} \sum_{j=1}^{n} \Delta \bar{t}_j = \frac{2 \bar{l}_F \beta}{c}, \tag{52}$$

where $\bar{l}_F = \lim_{n \to \infty} \sum_{j=1}^{n} d\bar{l}_j$. As explained above, $d\bar{l}_j$ is identical to the exact one $dl_j$. The sum $\bar{l}_F$ of differential lengths corresponds to the rest length of the fiber loop. Equation (52), which is the same as the time difference from ET [11], agrees with the experimental result. Using LT, one can find the exact time difference of the generalized Sagnac effect unless it is subject to the postulates of STR.

## 2. Numerical examples

The spatial vector obtained by the LT is not identical with that by the exact one. This subsection numerically examines how accurately the LT can find it. Given two constant velocity vectors, one can establish a 2-D spatial plane such that it includes them. Hence it can be assumed without loss of generality that $\boldsymbol{\beta}_i$ and $\boldsymbol{\beta}_j$ belong to the x-y plane in $S$. Then $\boldsymbol{\beta}_{ji}$ and $\boldsymbol{\beta}_{ij}$, which are linearly dependent on them, do so. We suppress the z-component so that the spatial and the velocity vectors are represented in 2-D spaces. To measure the accuracy of $\bar{\boldsymbol{x}}_{(j)}$ with respect to $\boldsymbol{x}_{(j)}$, we define a normalized error as follows:

$$e_x = \frac{\|\bar{\boldsymbol{x}}_{(j)} - \boldsymbol{x}_{(j)}\|}{\|\boldsymbol{x}_{(j)}\|},$$

where $\bar{\boldsymbol{x}}_{(j)}$ and $\boldsymbol{x}_{(j)}$ are obtained from Eqs. (28) and (6), respectively. To set the speed $\beta_i$ of the frame $S_i$, we consider the motion of the Earth. Our Solar System has been known to move at a speed of about 370 km/sec in the cosmic microwave background [33]. In the numerical calculation the speed of $S_i$ is set at 370 km/sec and $\boldsymbol{p}_{(i)} = [i, 1, 1]^T$ is used. The angle between $\boldsymbol{\beta}_i$ and $\boldsymbol{\beta}_j$ is denoted by $\phi$.

Figure 3 illustrates $e_x$ as a function of $\beta_j$ for $\phi = 5^o$, $15^o$, $45^o$, and $90^o$. The error increases with an increase in $\beta_j$. However it is very small even when the speed of $S_j$ is close to $c$. For example, when $\phi = 15^o$ and $\beta_j = 0.2$, $e_x$ is smaller than 3.3 x $10^{-5}$. As $\beta_j$ goes to zero, $T_L(\beta_{ji})$ approaches $T_G(\beta_j)$ so that $e_x$ tends to zero, which is shown in Fig. 3. Additionally one can see that the error increases with an increase in $\phi$.

In Fig. 4, the effect of $\phi$ on $e_x$ is shown when $\beta_j = 0.5$, 0.05, and 0.005. If $\phi$ is zero, $\beta_i$ and $\beta_j$ are collinear and the LT is the same as ET. As a result, $e_x$ reduces to zero. It is seen that as $\phi$ increases from zero, the error does so. When $\phi < 10^o$, the errors are shown to rapidly approach zero as $\phi$ decreases. Moreover we see that even if the speed of $S_j$ is close to $c$, the error is small irrespective of $\phi$. For example, when $\beta_j = 0.5$, it is less than 3.4 x $10^{-4}$ for every $\phi$.

## V. CONCLUSIONS

The STR has suggested the time relationship (30) and the Doppler formula (31). The predictions of STR have been verified through numerous experiments to test it. As seen in Eqs. (26a) and (27a), which are identical to Eqs. (30) and (31) from STR, PTs and Doppler shifts are exactly obtained by treating inertial frames as if they were isotropic, which may have led to the firm belief that STR has been experimentally verified. However, these experimental evidences prove PFT, rather than STR. If all inertial frames are equivalent and isotropic according to the postulates of STR, the left side of Eq. (33) should be equal to the right side, which is, however, mathematically infeasible. The Mocanu paradox results from the non-equality. Equations (26) and (27) have been derived from ET under the uniqueness of the isotropic frame. The reason why the experimental results of PTs and Doppler shifts are in agreement with the predictions of STR is because the first rows of $T_L(\beta_{ji})$ and $T_G(\beta_j, \beta_i)$ are identical, which leads LT to satisfy Eqs. (26a) and (27a).

Despite the infeasibility, LT must be a useful method to approach physics problems, as can be seen from the remarkable achievements that STR has shown so far. The exact physical quantities can be obtained from $T_G(\beta_j, \beta_i)$, which requires the absolute velocities $\beta_i$ and $\beta_j$, though. The LT allows us to make a transformation from $S_i$ to $S_j$ by using only the relative velocity $\beta_{ji}$, without the need for the absolute velocities. Though $T_L(\beta_{ji})$ is not equal to $T_G(\beta_j, \beta_i)$, some physical quantities, such as time intervals, spatial lengths, speeds of objects, and Doppler-shifted frequencies, can be exactly obtained from the former. It has been shown according to this fact that in the experiment of the generalized Sagnac effect, the speed of light with respect to PT and the travel times of light beams can be exactly calculated by utilizing LT. The analysis result via LT indicates that the speed of light is anisotropic in inertial frames. On the other hand, the components of the spatial vector, which are independent of synchronization schemes, are different from the exact ones. We have carried out numerical calculation to see how accurately the LT can find those. The numerical results show that the values from LT are, though not exact, very accurate even if the speed of $S_j$ is close to $c$.

The useful LT should be utilized, but with its exact meaning and its limitation. We cannot use it under the postulates of STR, which cause inconsistencies and paradoxes. The STR results in Eqs. (30) and (31) when $\boldsymbol{p}_{(i)}$ is directly Lorentz-transformed into $S_j$ as Eq. (28) with $\boldsymbol{\beta}_{ji}$ and $\boldsymbol{p}_{(i)}$ given correctly. If the coordinate vector in $S_j$ is obtained through an inertial frame $S_k$ so that it is calculated as $\overline{\boldsymbol{p}}_{(j)/k} = \boldsymbol{T}_L(\boldsymbol{\beta}_{jk}, \boldsymbol{\beta}_{ki})\boldsymbol{p}_{(i)}$ according to the postulates, even the time component of the resultant does not correspond to the one from the ET. We should use LT, keeping in mind that the isotropic frame is unique.

The experimental results of the Sagnac effect and the empirical evidences of the GPS indicate the anisotropy of the speed of light. Nonetheless, physics problems can be approached through the standard synchronization. Great insight of Einstein can be seen from the statements [34], "That light requires the same time to traverse the path A→M as for the path B→M is in reality neither a supposition nor a hypothesis about the physical nature of light, but a stipulation which I can make of my own freewill in order to arrive at a definition of simultaneity." Once the standard synchronization is introduced, the standard-synchronized frames may be considered equivalent so that the representation for physical laws can have the same form in each of them. As a matter of fact, as far as kinematics is concerned, the Galilean transformation conforms to the equivalence of inertial frames and so the preferred frame would not be revealed, though simultaneity is absolute. However, it is in disagreement with the experimental results of time dilation. Nature itself reveals the preferred reference frame.

# REFERENCES

[1] R. Mansouri and R. U. Sexl, Gen. Relativ. Gravit. **8** (7), 497 (1977).

[2] M. Kretzschmar, Z. Phys. A **342**, 463 (1992).

[3] C. M. Will, Phys. Rev. D **45**, 403 (1992).

[4] E. Riis *et al*., Phys. Rev. Lett, **60** (2), 81 (1988).

[5] G. Gwinner, Mod. Phys. Lett. A **20**, 791 (2005).

[6] R. Anderson, I. Vetharaniam, and G. E. Stedman, Phys. Rep. **295** (3–4), 93 (1998).

[7] D. W. MacArthur, Phys. Rev. D **33** (1), 33 (1986).

[8] B. Botermann *et al*., Phys. Rev. Lett. **113**, 120405 (2014).

[9]  S. Reinhardt *et al.*, Nat. Phys. **3**, 861 (2007).

[10] A. Einstein, Ann. Phys. Leipzig **17**, 891 (1905).

[11] Y.-H. Choi, Can. J. Phys. **95** (8), 761 (2017).

[12] Y.-H. Choi, Can. J. Phys. **94** (10), 1064 (2016).

[13] C. I. Mocanu, Arch. Elektrotech. **69** (2), 97 (1986).

[14] C. I. Mocanu, Found. Phys. Lett. **5**, 443 (1992).

[15] A. A. Ungar, Found. Phys. **19** (11), 1385 (1989).

[16] R. Wang, Y. Zheng, A. Yao, and D. Langley, Phys. Lett. A **312**, 7 (2003).

[17] R. Wang, Y. Zheng, and A. Yao, Phys. Rev. Lett., **93**, 143901 (2004).

[18] Y.-H. Choi, Eur. Phys. J. Plus **131** (9), 296 (2016).

[19] R. D. Klauber, Found. Phys. **37**, 198 (2007).

[20] G. Rizzi and M. L. Ruggiero, ed., *Relativity in Rotating Frames* (Kluwer Academic, Dordrecht, The Netherlands, 2004).

[21] A. Ori and J. E. Avron, Phys. Rev. A **94** (6), 063837 (2016).

[22] J. P. Costella, B. H. J. McKellar, and A. A. Rawlinson, Am. J. Phys. **65** (8) 837 (2001).

[23] K. O'Donnell and M. Visser, Eur. J. Phys. **32** (7), 1033 (2011).

[24] K. Rebilas, Eur. J. Phys. **34** (3), L55 (2013).

[25] F. Selleri, Found. Phys. Lett. **18** (4), 325 (2005).

[26] Y.-H. Choi, Phys. Essays **30** (4), 364 (2017).

[27] G. Rizzi, M. L. Ruggiero, and A. Serafini, Found. Phys. **34** (12), 1835 (2004).

[28] A. L. Kholmetskii and T. Yarman, Can. J. Phys. **93** (5), 503 (2015).

[29] R. P. Feynman, R. B. Leighton, and M. Sands, *The Feynman Lectures on Physics* (Addison-Wesley, Massachusetts, Palo Alto and London, 1964).

[30] N. Ashby, Living Rev. Relativ. **6**, 1 (2003).

[31] Y.-H. Choi, Phys. Essays **29** (3), 440 (2016).

[32] J. C. Hafele and R. E. Keating, Science **177**, 166 (1972).

[33] G. F. Smoot, Rev. Mod. Phys. **79** (4), 1349 (2007).

[34] A. Einstein, *Relativity*: *The Special and the General Theory* (Crown, New York, 1952).

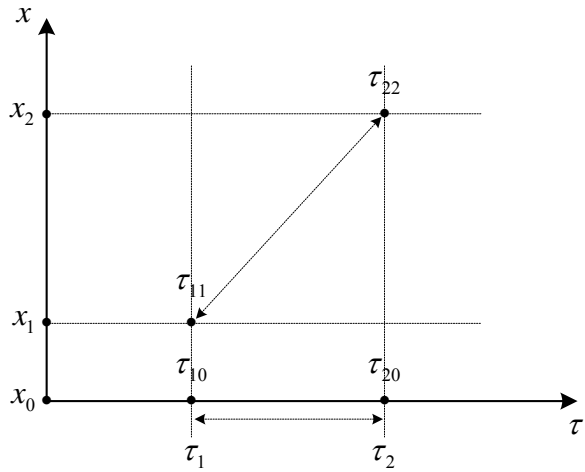

Fig. 1. Time interval at the same place and time interval between different places, which are equal in the isotropic frame $S$.

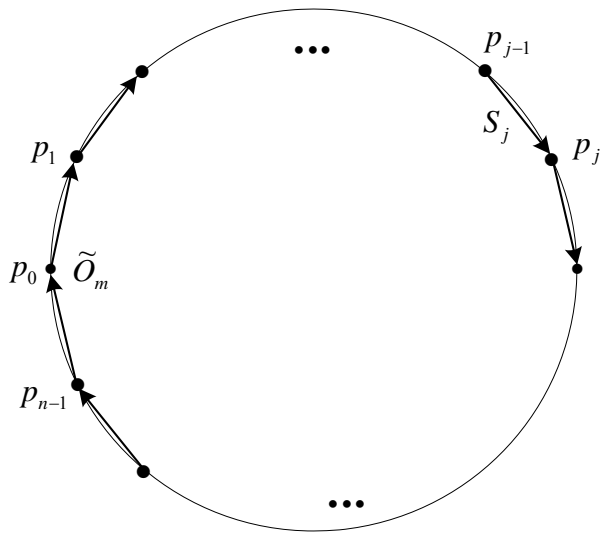

Fig. 2. Linearized description of a circle by $n$ line segments.

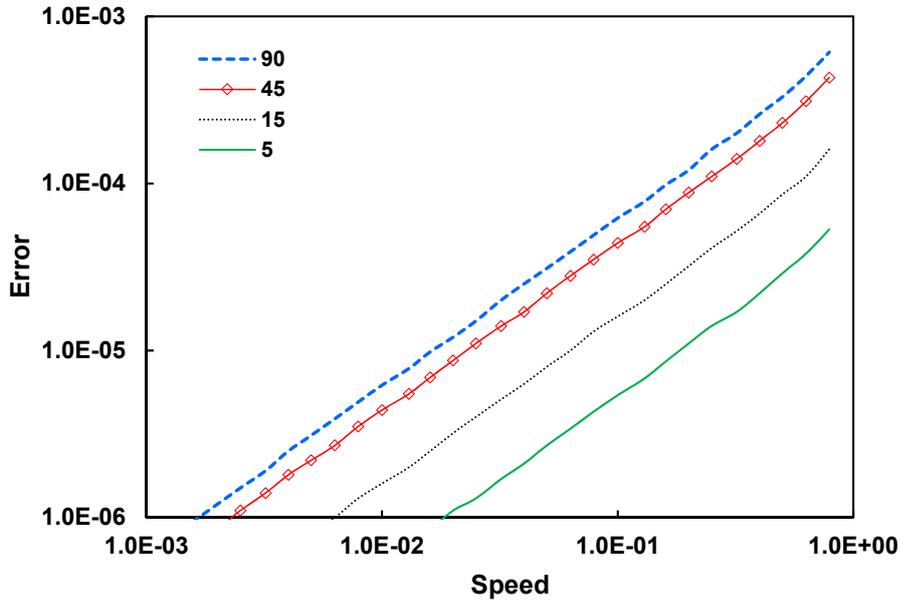

Fig. 3. Normalized errors versus $\beta_j$ for $\phi = 5°$, $15°$, $45°$ and $90°$.

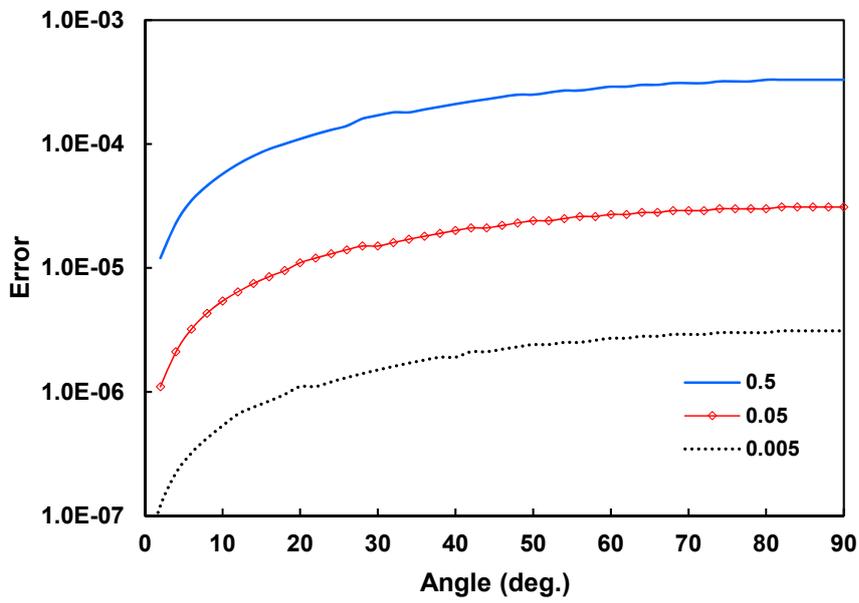

Fig. 4. Normalized errors versus $\phi$ for $\beta_j = 0.5$, $0.05$, and $0.005$.